\documentclass[fleqn,10pt]{SelfArx} 

\definecolor{color1}{RGB}{0,0,90} 
\definecolor{color2}{RGB}{0,20,20} 

\usepackage{hyperref} 
\hypersetup{hidelinks,colorlinks,breaklinks=true,urlcolor=color2,citecolor=color1,linkcolor=color1,bookmarksopen=false,pdftitle={Title},pdfauthor={Author},urlcolor=blue}

\usepackage{graphicx}
\usepackage[square]{natbib}
\usepackage{amsmath}
\usepackage{multirow}
\usepackage{subfigure}
\usepackage{siunitx}
\usepackage{geometry}
\usepackage{pdflscape}

\newcommand{\n}[1]{\mathrm{#1}}
\newcommand{\etapv}{\eta_\mathrm{PV,0}}

\JournalInfo{Published in Solar Energy, Vol. 120, 187-194, 2015} 
\Archive{\href{http://dx.doi.org/10.1016/j.solener.2015.07.035}{DOI: 10.1016/j.solener.2015.07.035}} 

\PaperTitle{The performance of a combined solar photovoltaic (PV) and thermoelectric generator (TEG) system} 

\Authors{R. Bj\o{}rk, K. K. Nielsen} 
\affiliation{\textit{Department of Energy Conversion and Storage, Technical University of Denmark - DTU, Frederiksborgvej 399, DK-4000 Roskilde, Denmark}} 
\affiliation{*\textbf{Corresponding author}: rabj@dtu.dk} 

\newcommand{\keywordname}{Keywords} 

\Abstract{The performance of a combined solar photovoltaic (PV) and thermoelectric generator (TEG) system is examined using an analytical model for four different types of commercial PVs and a commercial bismuth telluride TEG. The TEG is applied directly on the back of the PV, so that the two devices have the same temperature. The PVs considered are crystalline Si (c-Si), amorphous Si (a-Si), copper indium gallium (di)selenide (CIGS) and cadmium telluride (CdTe) cells. The degradation of PV performance with temperature is shown to dominate the increase in power produced by the TEG, due to the low efficiency of the TEG. For c-Si, CIGS and CdTe PV cells the combined system produces a lower power and has a lower efficiency than the PV alone, whereas for an a-Si cell the total system performance may be slightly increased by the TEG.}

\begin{document}

\flushbottom 

\maketitle 


\thispagestyle{empty} 

\section{Introduction}
Converting the energy from the Sun directly to electricity in an efficient way is of great interest. Photovoltaic devices (PV) can directly convert parts of the solar spectrum, but a significant part is absorbed as heat. In order to remedy this, a number of combined photovoltaic and heat recovery systems has been proposed recently. The most simple of these convert the heat energy directly to electrical energy using a thermoelectric generator (TEG). The latter utilizes the Seebeck effect to convert heat directly into electrical energy through the movement of charge carriers induced by a temperature span across the TEG.

A number of such combined systems has been studied recently \cite{Sundarraj_2014}. \citet{Luque_1999} considered a general hybrid PV and thermal system and showed that such systems may generally have a higher efficiency than single PV systems. A more specific combined PV + TEG system that uses a wavelength separating device to separate the incoming solar radiation into two parts, up to 800 nm for the PV and longer wavelengths to the TEG, has also been proposed \cite{Zhang_2005,Kraemer_2008}. An experimental realization, using a hot mirror and a near-infrared focusing lens was made by \citet{Mizoshiri_2012}, who observed an increase in total open circuit voltage of 1.3 \% compared to that of the PV alone. A similar system was also modeled where it was found that the TEGs contributed about 10 \% of the output power in the hybrid system\cite{Ju_2012}.

Coupled PV + TEG systems have also been considered solely from a modeling basis \cite{Sark_2011,Kiflemariam_2014,Liao_2014,Zhang_2014,Attivissimo_2015,Wu_2015}. Specifically, \citet{Vorobiev_2006} considered a combined PV and TEG system, with possible concentration of the heat from the PV to the TEG, using a modeling approach. Concentration of the incoming solar radiation was shown to lead to an increase in efficiency of 5-10 \%, albeit with highly efficient TEGs, whereas with no concentration the efficiency of the combined system is lower than that of the PV alone. This is also found by \citet{Lin_2014}, using an iterative numerical scheme which accounts for the absorbed solar spectrum. \citet{Attivissimo_2015} find similar values, ranging from 1-16 \%, for no solar concentration, but depending on the geographical location of the system. Similarly, \citet{Zhang_2014} found an increase in efficiency of 1-30\%, for conventional PV systems, while \citet{Xu_2014} found an increase of 8\% for a combined PV + TEG system. \citet{Najafi_2013} modeled a combined PV + TEG system where the TEG modules were attached to the back side of the PV collector. For 2.8 Suns solar irradiance the considered PV + TEG system produced 145 W by the PV panel and 4.4 W by the TEG modules. For another case, the power generated by the TEGs were found to be 1.84 \% of the total generated electricity by the PV panel. A small experimental system by \citet{Fisac_2014} also finds merely a small increase in efficiency by adding a TEG to a PV.

However, a different class of results exists. Experimental realizations of small combined PV and TEG systems have reported a significant increase in performance compared to that of a PV alone. Originally, \citet{Wang_2011} reported an experimental realization of a combined dyesensitized solar cell (DSSC) PV and TEG system. The reported efficiency of the PV alone in the combined PV + TEG system is 9.39 \% whereas the total efficiency of the PV + TEG is 13.8 \%, although the temperature span of the TEG was reported to be just 6.2 $^\circ$C. This means that the efficiency of the TEG is at least 4.87 \%, when all excess heat passes through the TEG. This is the same value of efficiency as the reported efficiency of commercial TEG modules at a temperature span of 200 $^\circ$C, for which the efficiency is $\approx$5 \% \cite{Marlow_2014}. The commercial module used in \citet{Wang_2011}, a Micropelt MPG-D602, has approximately a maximum power produced of 0.25 mW at a temperature span of 5 K and correspondingly a total heat flux of 0.35 W thus resulting in an efficiency of approximately 0.07 \% based on the official documentation from Micropelt \cite{Micropelt}, a number drastically different than the $\approx5\%$ reported. Furthermore, the Carnot efficiency, $\eta_\mathrm{Carnot}=\frac{\Delta T}{T_\mathrm{hot}}$ for a room-temperature TEG with a temperature span of 6.2 $^\circ$C is 2.1 \% and so the reported results clearly appear unphysical.

Similar results were obtained by \citet{Hsueh_2014}, who studied an experimental realization of a coupled thin film CIGS solar cell with a TEG. Here the efficiency of the PV alone in the combined PV + TEG system was 16.5 \% whereas the total efficiency of the PV + TEG was 22.02 \%, although the temperature span of the TEG was reported to be just 11.6 $^\circ$C. This suggests a minimum efficiency of the TEG of 6.61 \%, at a very small temperature span. \citet{Park_2013} also constructed a PV + TEG system and observed an increase in efficiency from 12.5 \% to 16.3 \% , corresponding to an increase in power output by 30\%, although the temperature gradient across the TE device was 15 $^\circ$C. Given the application of 127 legs with a cross sectional area of $6.4\cdot10^{-3}\ \mathrm{cm}^2$ and length of 0.05 cm \cite{Park_2013} results in a total thermal resistance of 6.2 K/W. At a temperature span of 15 $^\circ$C this results in a hot side heat flux of 2.4 W. At a claimed power production of 30 mW \cite{Park_2013} this results in a TEG module efficiency of 1.3 \%, which is substantially less than the reported 6.61 \%. Again, the Carnot efficiency in this case is 3.9 \%, which as above renders the results unphysical.

\citet{Zhang_2013} combined a polymer solar cell with a TEG and showed an increased output power of 46.6\%, even though the temperature span was only 9.5 $^\circ$C. The same is observed by \citet{Deng_2013}, who for a combined PV + TEG system had a conversion efficiency of 4.55 \% for the PV alone, which more than doubled once a TEG was added to the system, although the temperature span was only 7 $^\circ$C. However, in this last case a large concentrator was used to collect the heat. The Carnot efficiency at these operating conditions is 3.2 \% once again rendering the results unphysical. Explaining these results in terms of the current understanding of thermoelectric generators is not possible, as the efficiency is simply too high compared to well established reported material properties as well as the known performance of commercial TEG devices.

\citet{Sark_2011} modeled an idealized case of combining a PV and TEG assuming state-of-the-art TEG performance (with a thermoelectric figure of merit $z$ value of 0.004 K$^{-1}$). It was found that at a temperature span of 60 $^\circ$C the maximum efficiency for the theoretically optimal TEG module would be 3.2 \%. This is a  significantly lower TEG efficiency than those stated in the papers mentioned above (where the temperature span ranged from 5 to 20 $^\circ$C).  However, only a multi-crystalline silicon PV was considered.

In this paper, we derive the efficiency of such a coupled PV and TEG system, with the TEG sitting directly on the back of the PV, using an analytical approach. The efficiency of such a system cannot be considered as a single stand-alone system and thus its power generation characteristics are not known. Clearly determining the efficiency of such a combined system will also aid in the discussion of the above-mentioned results. We note that the cold side of the TEG is assumed to be perfectly cooled to the reference temperature (25 $^\circ$C in the remainder of this paper). In the analysis conducted below, no considerations on how to provide this cooling are taken into account. The analysis presented differs from previous work in being a pure analytical model, e.g. unlike the iterative model of \citet{Lin_2014}, and  relaying purely on experimentally determined TEG performance, e.g. unlike \citet{Sark_2011}.

\section{The studied system}
\begin{figure}[!t]
  \centering
  \includegraphics[width=1\columnwidth]{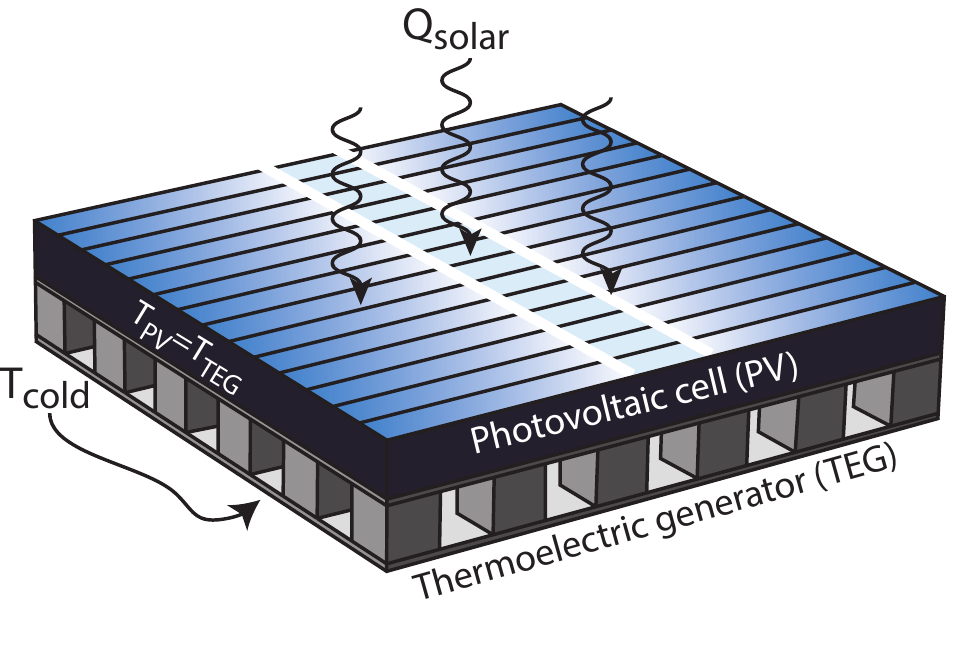}
  \caption{An illustration of the system considered. The incoming solar radiation hits the PV, which heats up from parts of the radiation. This heat is transferred through the TEG, to the cold side.}
  \label{Fig_Illustration_PV_TEG}
\end{figure}

We consider a combined PV and TEG system, where the TEG is mounted directly on the back of the PV. The hot side temperature of the TEG is thus equal to the temperature of the PV, $T_\n{PV}=T_\n{TEG, hot}$. The system is shined upon by the Sun, with an arbitrary concentration of the light. The studied setup is illustrated in Fig. \ref{Fig_Illustration_PV_TEG}.

In order to derive the power generated and the corresponding efficiency of the coupled system, the properties of the PV and TEG as a function of temperature must be known. This is described in the following. It is assumed that the properties of the PV do not change with solar concentration, but only with temperature.

\subsection{Properties of PVs as a function of temperature}
We consider four different kinds of PVs, namely crystalline Si (c-Si), amorphous Si (a-Si), copper indium gallium (di)selenide (CIGS) and cadmium telluride (CdTe). The relevant performance characteristics of the different PVs are given in Table \ref{Table_1}. The degradation in PV efficiency with temperature is given by
\begin{eqnarray}
\eta_\n{PV} = \eta_{T_\n{ref}}\left(1+\beta(T_\n{PV,0}-T)\right)\label{eq_eff_PV}
\end{eqnarray}
where $T_\n{ref}$ is the reference temperature, typically 25 $^\circ$C, and $\beta$ is the temperature coefficient \cite{Skoplaki_2009}. The values for the temperature coefficient for the different types of PVs is reported in Table \ref{Table_1}, and are in agreement with data reported elsewhere in literature \cite{Emery_1996, Mohring_2008, Makrides_2009, Makrides_2012}.

We assume that the photons provided from the Sun are either converted directly into electrical energy, with fraction $\eta_{PV}$, converted into heat, $\eta_{T}$, or not absorbed, $\eta_\mathrm{non}$, such that
\begin{eqnarray}\label{sum_eff}
\eta_\n{PV}+\eta_\n{T}+\eta_\n{non} = 100\%
\end{eqnarray}
The fraction of non-absorbed photons is taken to be constant and the values for c-Si, CIGS, CdTe and a-Si PVs are 16\%, 18\%, 37\% and 47\%, respectively \cite{Lorenzi_2014}, as also given in Table \ref{Table_1}. The optical losses are not included, but are assumed to be small \cite{Lorenzi_2014}. These data fit well with measurements of the total absorption factor of a typical encapsulated c-Si PV, which can be as high as 90.5\% \cite{Santbergen_2008}.

\subsection{Properties of TEGs as a function of temperature}
The efficiency of a TEG is generally a nonlinear function of temperature since the thermoelectric material properties vary nonlinearly with temperature. In the case when the material properties do not depend on temperature an analytic expression for the efficiency can be derived. This depends on both the hot and cold side temperature and the temperature span. However, this assumption is not applicable for actual thermoelectric materials.

For hot side temperatures below 250 $^\circ$C, bismuth telluride, Bi$_2$Te$_3$ (BiTe), is the thermoelectric material with the highest efficiency, by a large margin. The thermoelectric figure of merit, $zT$, is 1.5 for $p$-type and 0.8 for $n$-type Bi$_2$Te$_3$ at 100 $^\circ$C \cite{Ma_2008,Kim_2012}. For practical applications, a number of commercial Bi$_2$Te$_3$ modules exist. For the analysis presented here, we consider one such commercial module, TG12-4, from Marlow Industries, Inc. However, the performance of this device is more of less identical to other Bi$_2$Te$_3$ modules from various manufacturers.

\begin{figure}[!t]
  \centering
  \includegraphics[width=1\columnwidth]{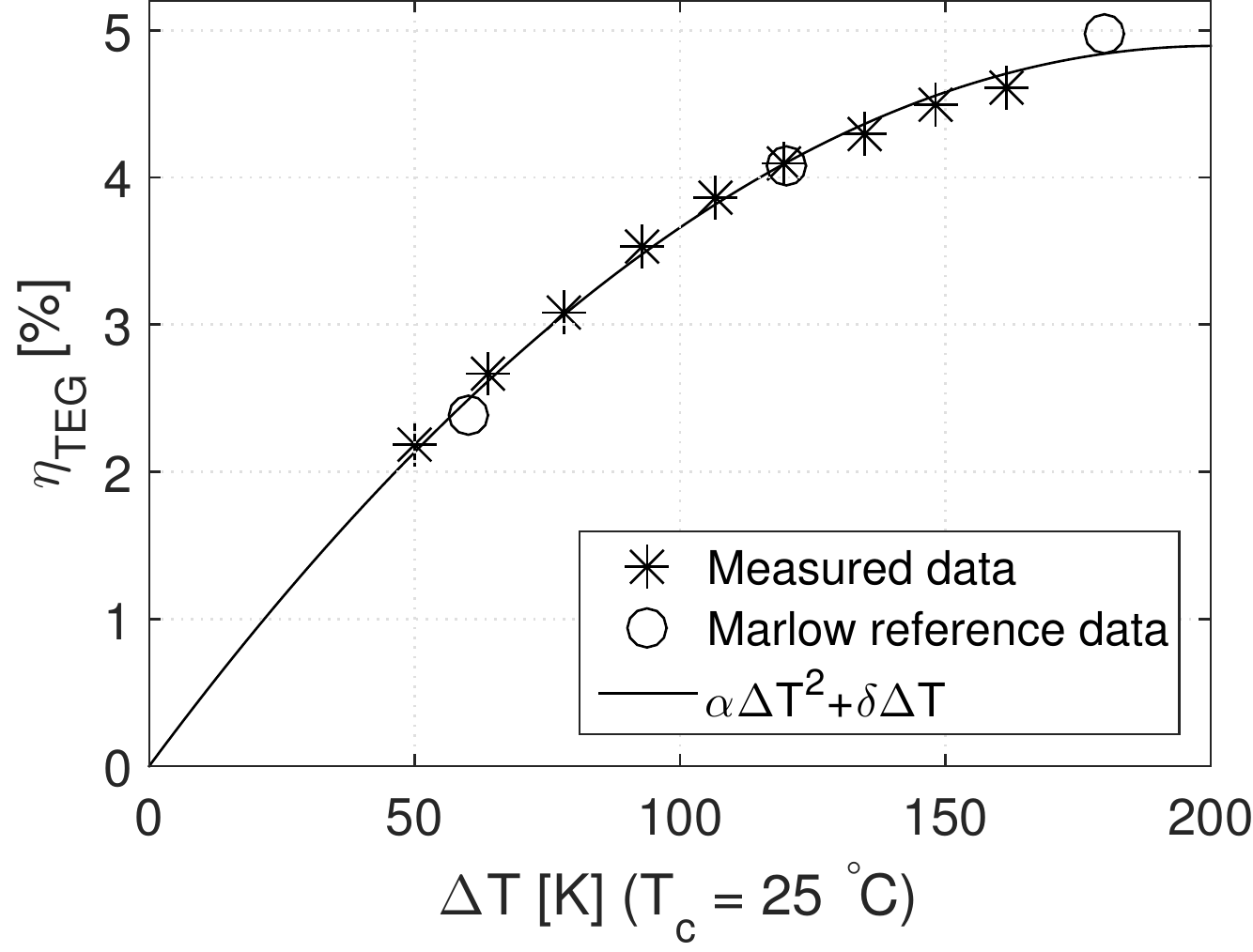}
  \caption{The measured efficiency of a commercial thermoelectric generator as a function of temperature span, for a cold side temperature of 25 $^\circ$C. Reference data is also shown. A function of the form $\eta = \alpha{}\Delta{}T^2+\delta\Delta{}T$, has been fitted to the data.}
  \label{Fig_Marlow_eff_Tspan}
\end{figure}

The efficiency as a function of temperature for this module was measured using an in-house thermoelectric module tester \cite{Hung_2015}, and the results are shown in Fig. \ref{Fig_Marlow_eff_Tspan} along with datasheet reference values. In order to use these data for the analytic model developed, a second degree polynomial in $\Delta T$ was fitted to the data. The fit has an R$^2$ value of 0.9996, indicating a good fit to the data. For the Marlow BiTe commercial module the efficiency as a function of temperature span is thus given by
\begin{eqnarray}
\eta_\n{TEG} = \alpha{}\Delta{}T^2+\delta\Delta{}T\label{eq_eff_TEG}
\end{eqnarray}
where $\alpha{} = -1.21(5)*10^{-6}\ \mathrm{K}^{-2}$ and $\delta = 4.87(7)*10^{-4}\ \mathrm{K}^{-1}$ for a cold side temperature of 25 $^\circ$C.

\subsection{Results}
We now consider the combined PV + TEG system. The heat flux available at the hot side of the TEG is assumed to be what remains of the absorbed radiation of the PV power production:
\begin{equation}
Q_\mathrm{TEG}(T)=Q_\mathrm{solar}(1-\eta_{PV}(T)-\eta_\mathrm{non}).\label{eq_q_TEG}
\end{equation}
The total electrical power produced is the sum of the power produced by the PV and by the TEG:
\begin{equation}
P_\mathrm{tot}(T)=Q_\mathrm{solar}\eta_{PV}(T)+Q_\mathrm{TEG}(T)\eta_\mathrm{TEG}(T).\label{eq_Ptot}
\end{equation}
It is straightforward, although tedious, to find the extrema of Eq. \ref{eq_Ptot} as a function of temperature by combining Eqs. (\ref{eq_eff_PV}), (\ref{eq_eff_TEG}) and (\ref{eq_q_TEG}); see Appendix \ref{sec_app_max_eff}. The maximum total power produced by the combined PV and TEG system can then be obtained.

\begin{figure*}
\begin{center}
\subfigure[c-Si]{\includegraphics[width=0.47\textwidth]{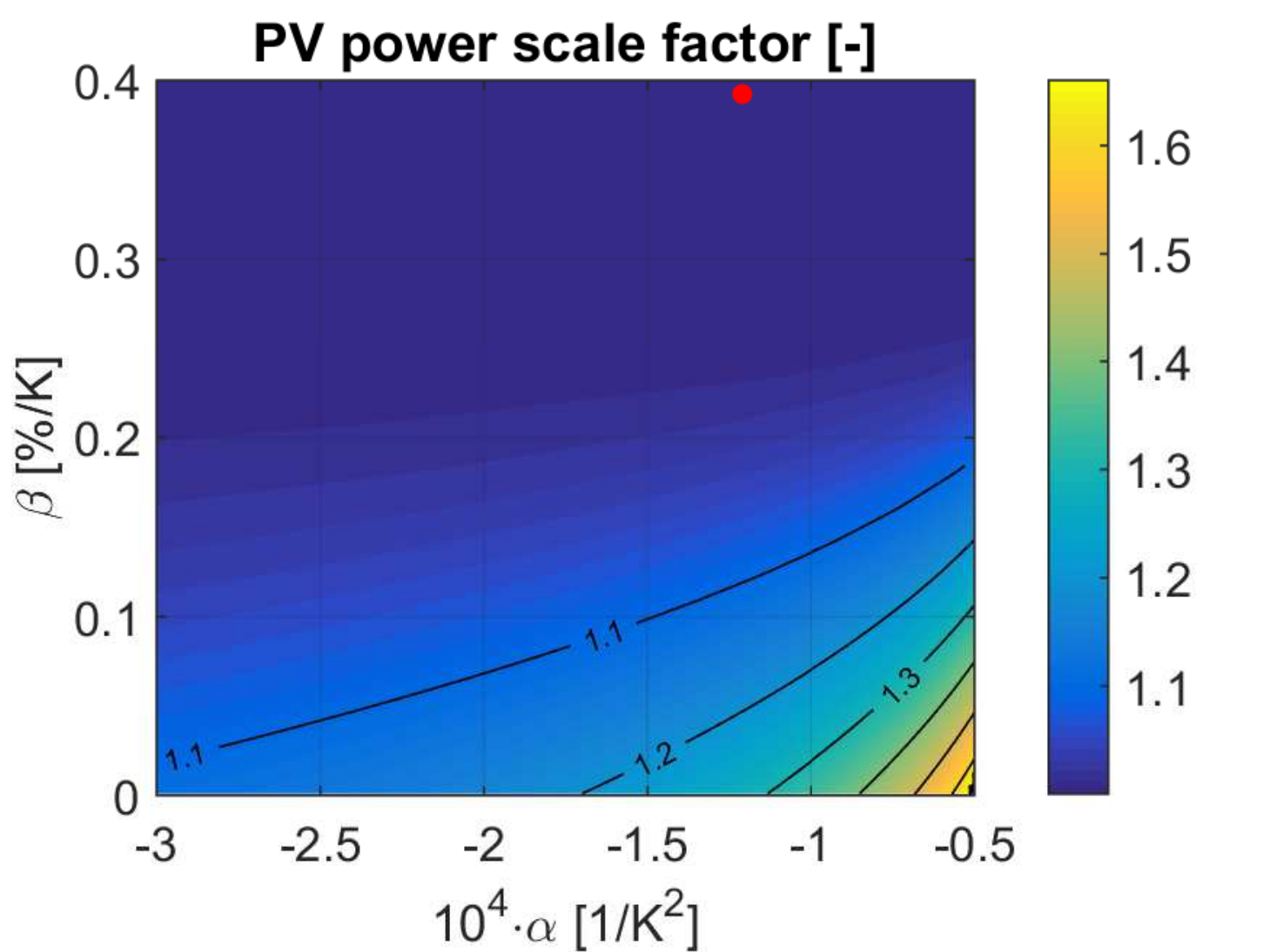}}
\subfigure[a-Si]{\includegraphics[width=0.47\textwidth]{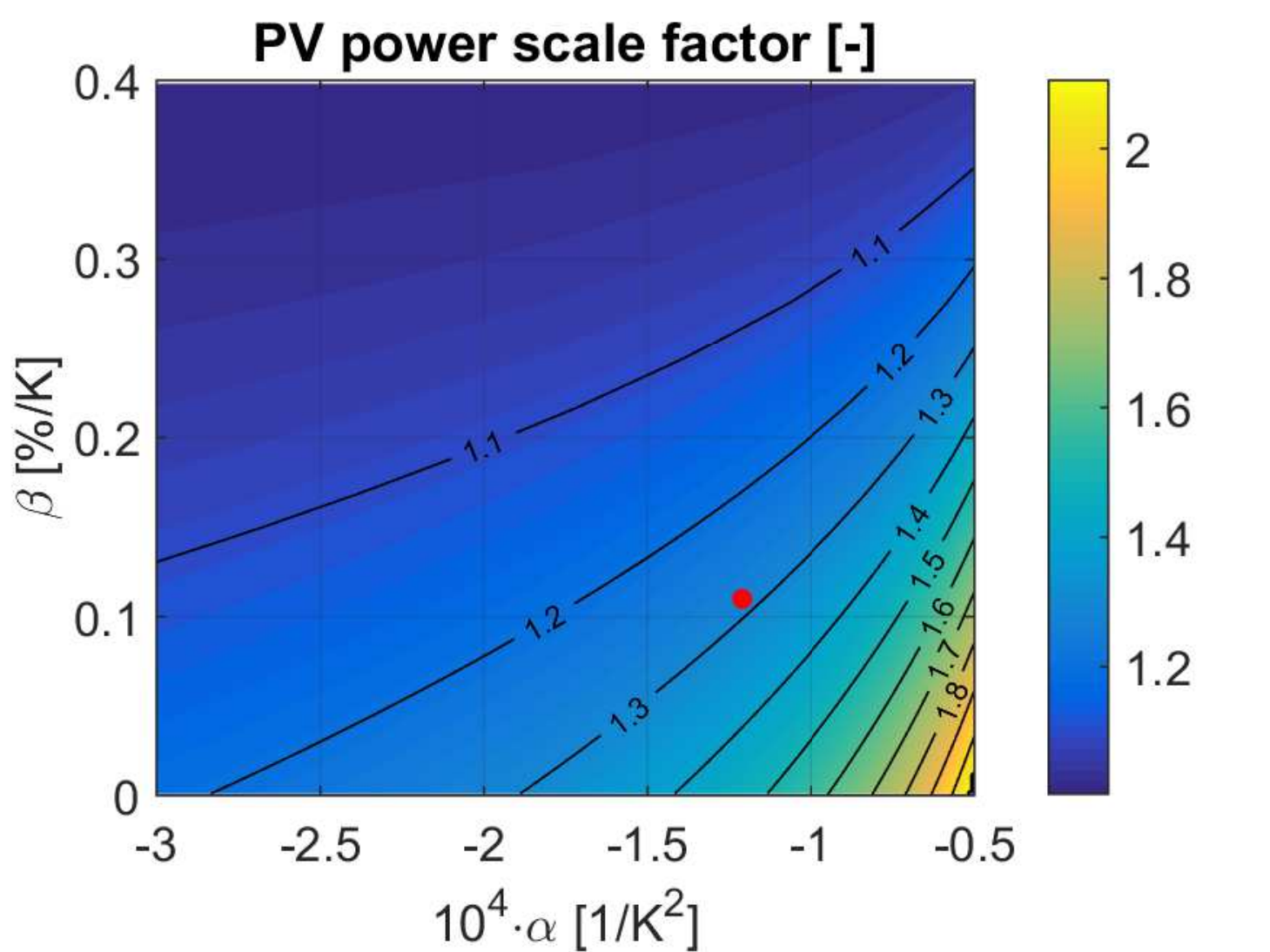}}
\subfigure[CIGS]{\includegraphics[width=0.47\textwidth]{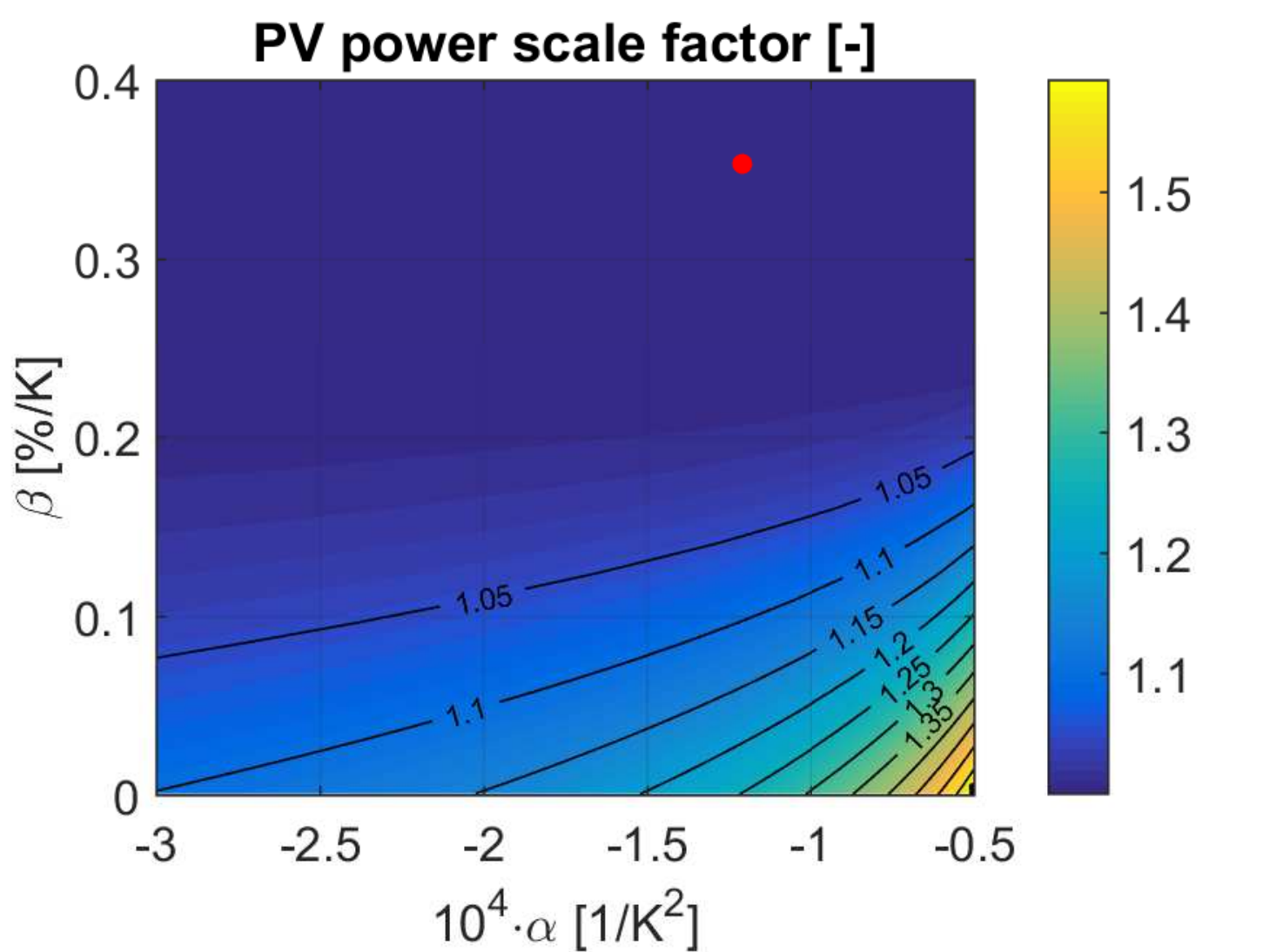}}
\subfigure[CdTe]{\includegraphics[width=0.47\textwidth]{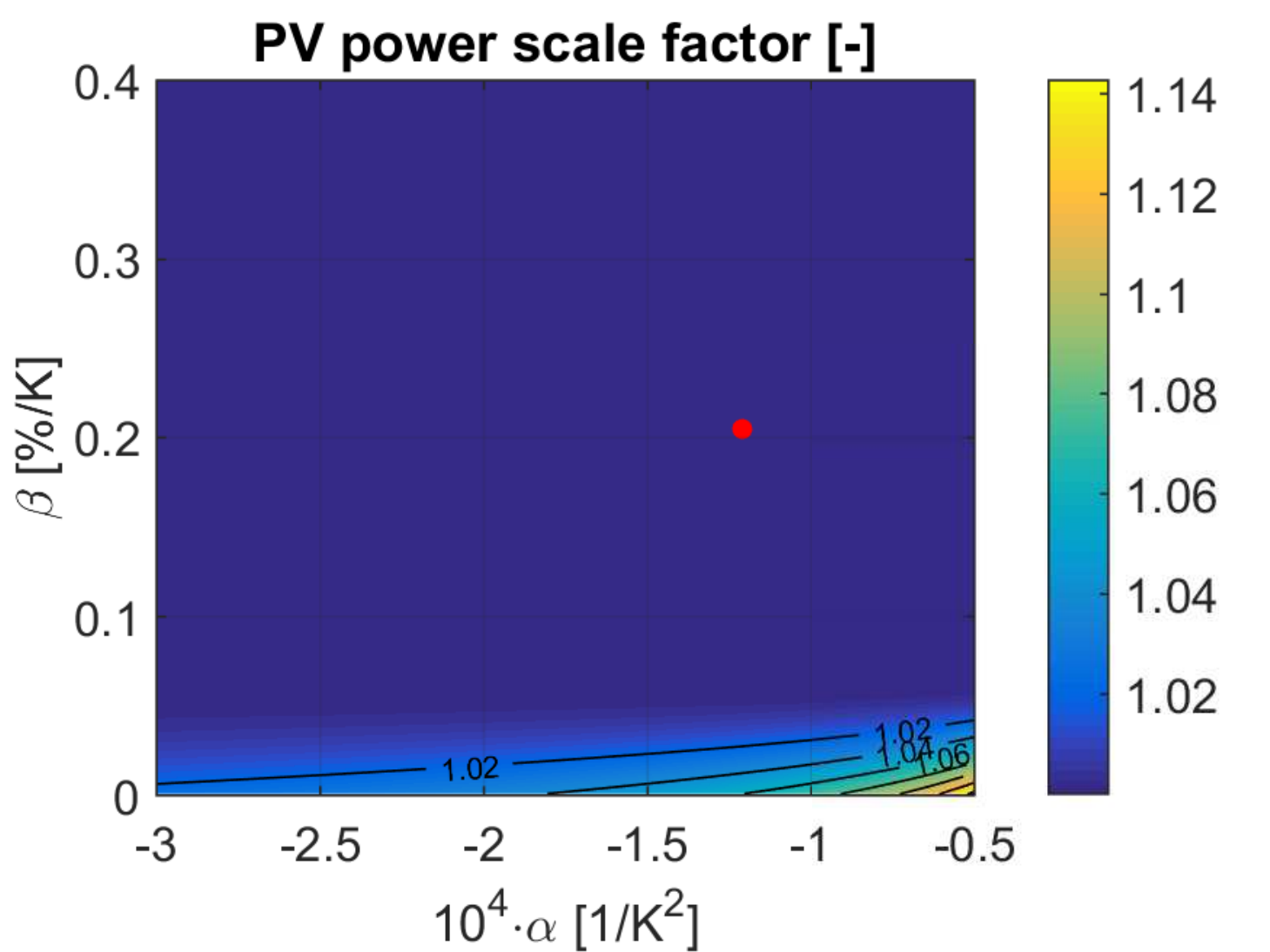}}
\end{center}
\caption{The maximum power scale factor, defined in Eq. (\ref{eq_MaxPowerScale}), as a function of the thermal degradation coefficients for the PV ($\beta$) and the TEG ($\alpha$). The red dot in each plots indicate the actual set of PV and TEG parameters from Table \ref{Table_1}. The difference between the figures are the values of $\eta_\n{NON}$ and $\eta_\n{PV,0}$.}\label{Fig_Power_PV_TEG_alpha_beta_maps}
\end{figure*}

\begin{figure*}
\begin{center}
\subfigure[c-Si]{\includegraphics[width=0.47\textwidth]{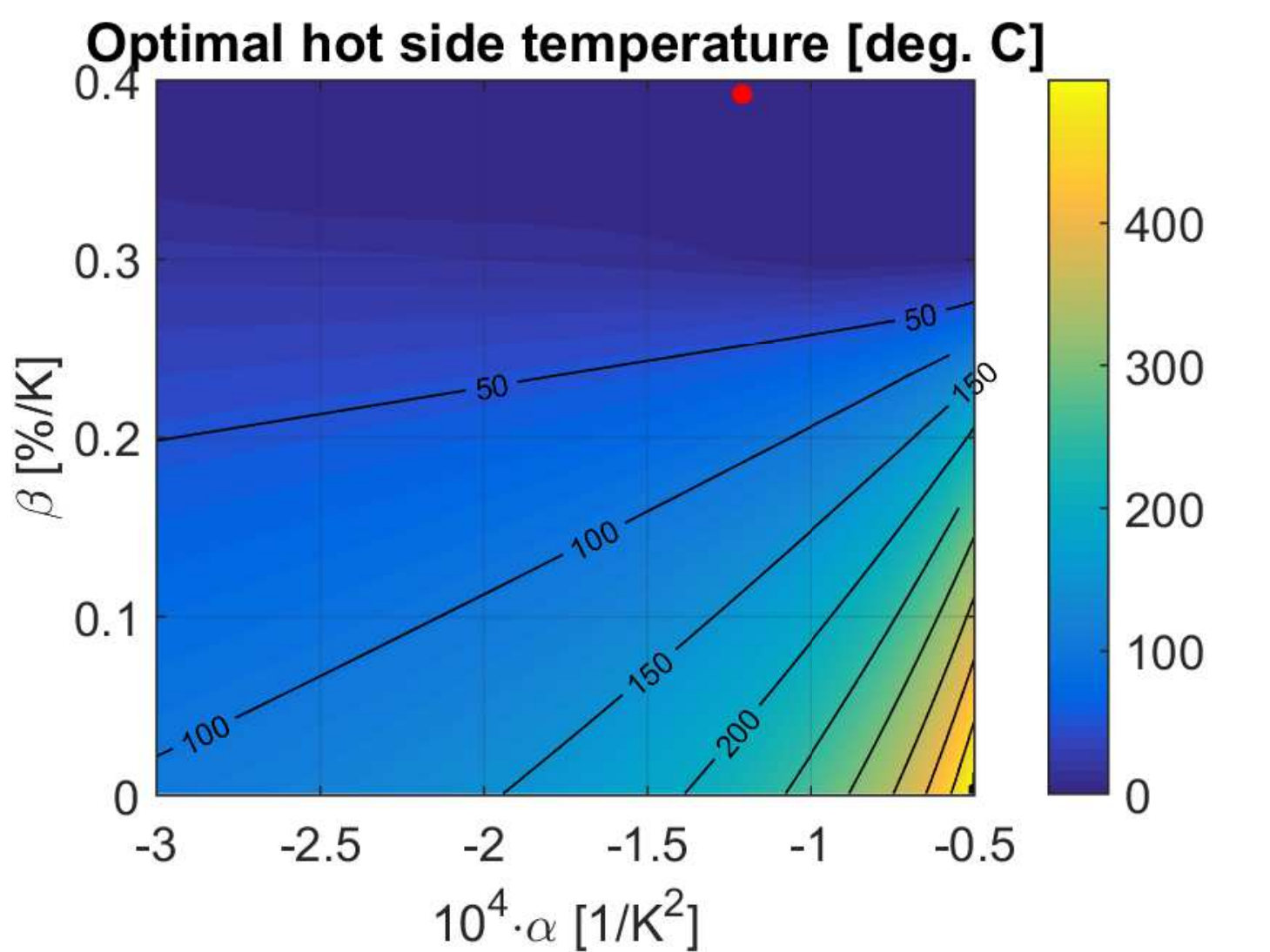}}
\subfigure[a-Si]{\includegraphics[width=0.47\textwidth]{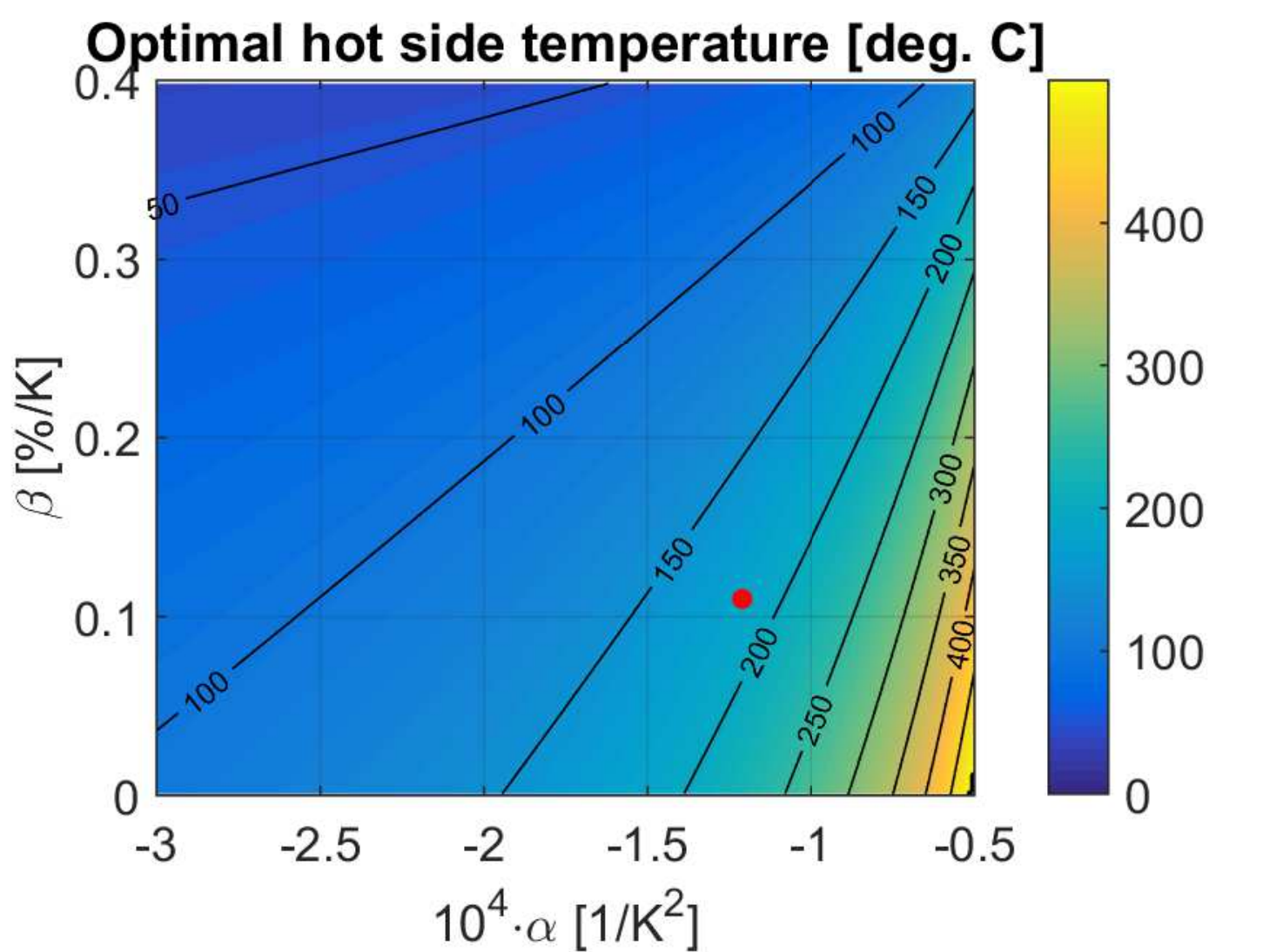}}
\subfigure[CIGS]{\includegraphics[width=0.47\textwidth]{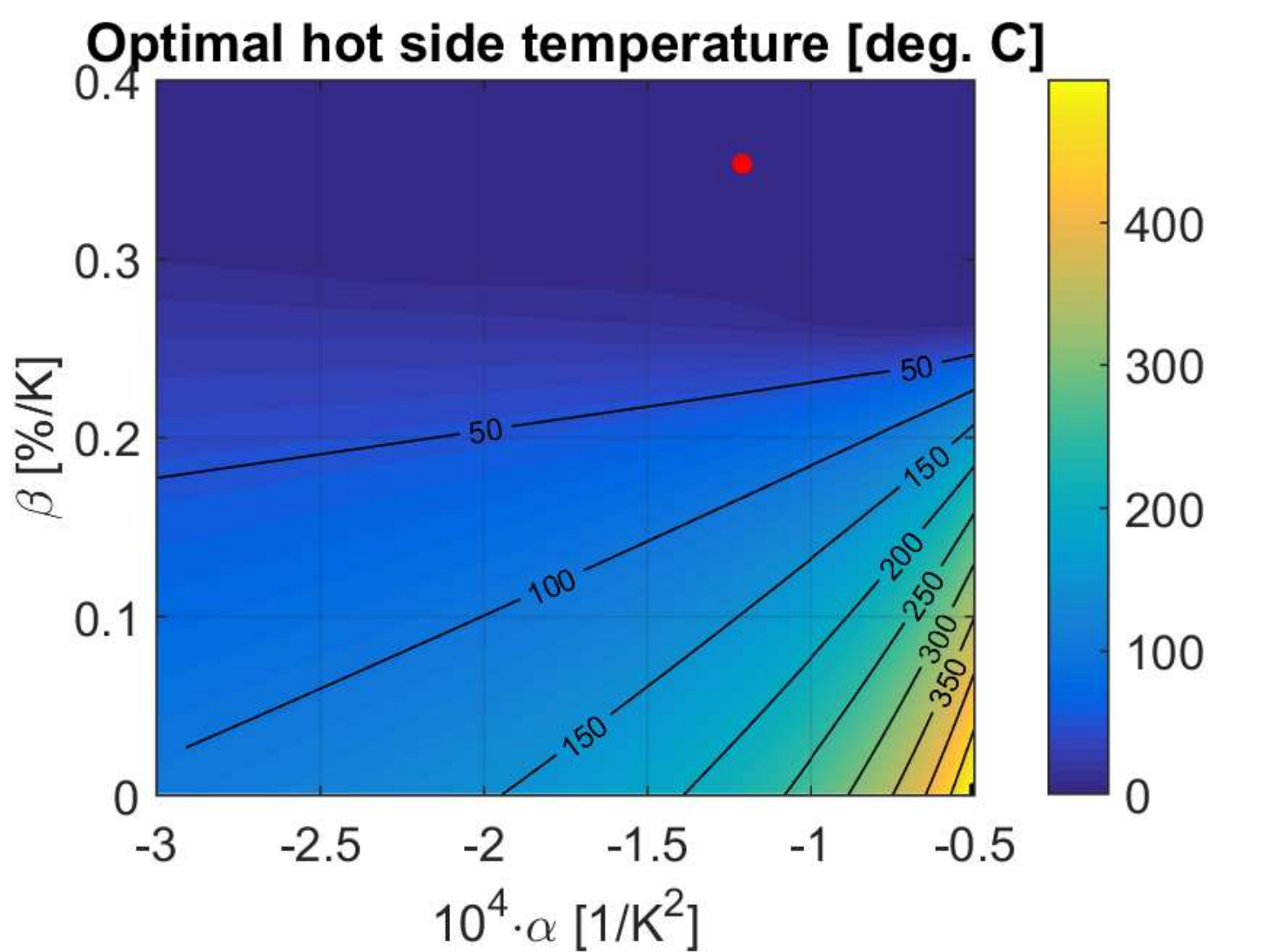}}
\subfigure[CdTe]{\includegraphics[width=0.47\textwidth]{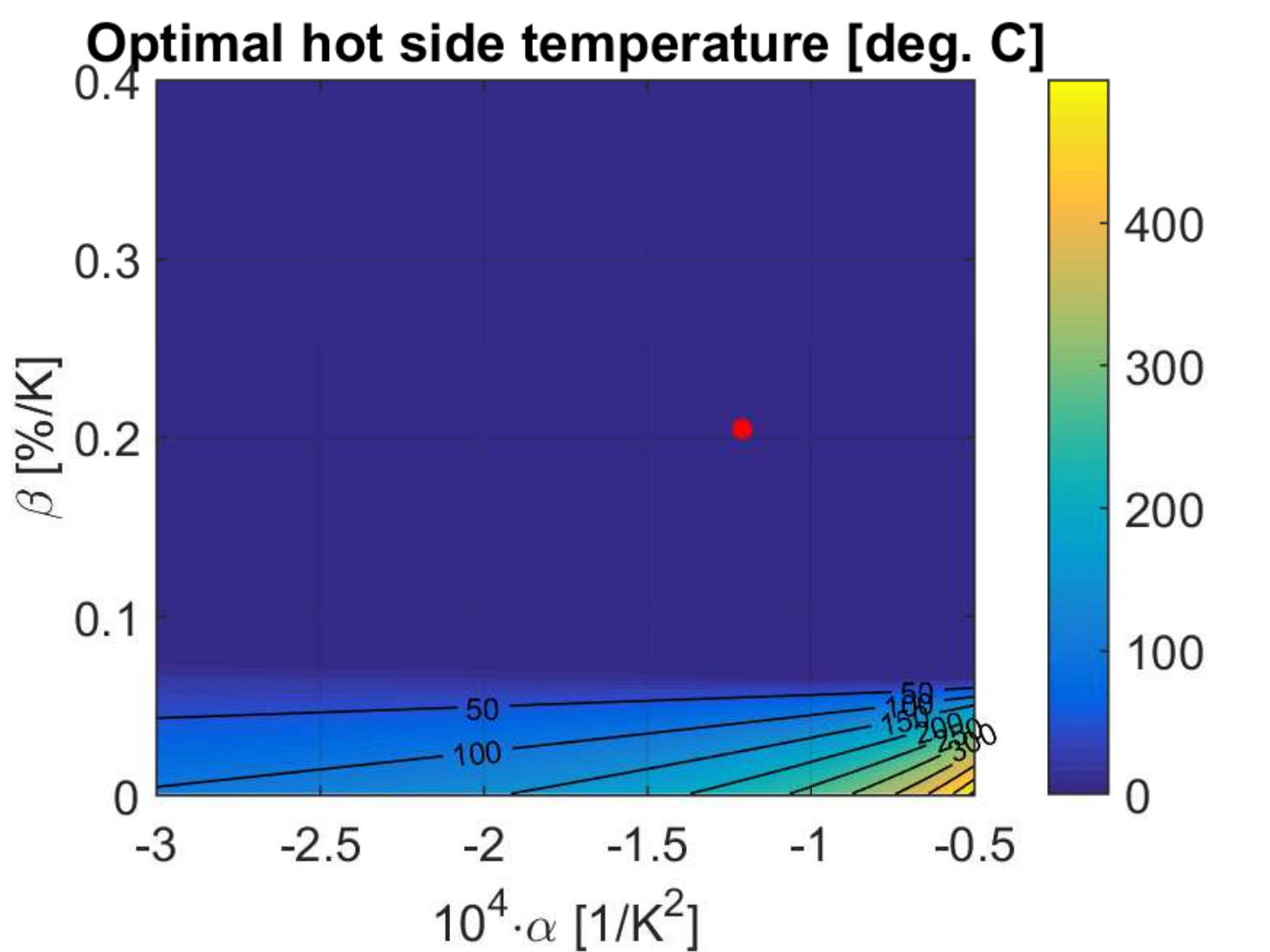}}
\end{center}
\caption{The optimal temperature (where the power scale factor is at a maximum) as a function of $\beta$ and $\alpha$. The red dot in each plot indicates the parameter configuration of the actual PV + TEG combination.}\label{Fig_Tmax_PV_TEG_alpha_beta_maps}
\end{figure*}

Initially, we consider the performance as a function of the thermal degradation coefficient of the PV, $\beta$, and the quadratic constant of the TEG, $\alpha$. We consider the performance in terms of the increase in power, compared to the power produced by a PV alone, i.e.:
\begin{equation}
\gamma = \frac{\mathrm{max}(P_\mathrm{tot})}{P_\mathrm{PV only}},\label{eq_MaxPowerScale}
\end{equation}
This is plotted in Fig. \ref{Fig_Power_PV_TEG_alpha_beta_maps} with the corresponding optimum temperature maps plotted in Fig. \ref{Fig_Tmax_PV_TEG_alpha_beta_maps}.

\clearpage

\begin{landscape}
\begin{table}[p]
\begin{center}
\caption{The relevant absorber and temperature properties of the considered PVs. The values given for CdTe are theoretical maximum values. The CIGS module data are from commercial modules, which seems to be accurate in tests \cite{Dunn_2012}. The values for c-Si are an average of values from \citet{Skoplaki_2009}.}
\begin{tabular}{|l|rr|rr|rr|}\label{Table_1}
PV & $\eta_\n{non}$ [\%] & Ref. & $\eta_\n{PV,0}$ [\%] & Ref. & $\beta$ [\% K$^{-1}$] & Ref. \\
\hline
c-Si & 16 & \cite{Lorenzi_2014} & 12.4 & \cite{Skoplaki_2009} & 0.392 & \cite{Skoplaki_2009} \\ 
a-Si & 47 & \cite{Lorenzi_2014} & 5.0 & \cite{Skoplaki_2009} & 0.110 & \cite{Skoplaki_2009}\\
\multirow{ 2}{*}{CIGS} & \multirow{ 2}{*}{18} & \multirow{ 2}{*}{\cite{Lorenzi_2014}} & \multirow{ 2}{*}{13.3} & \cite{TSMC_2015} & \multirow{ 2}{*}{0.353} & \cite{TSMC_2015} \\
 &  &  &  & \cite{Hulk_2015} &  & \cite{Hulk_2015} \\
 &  &  &  & \cite{Solibro_2015} &  & \cite{Solibro_2015} \\
CdTe & 37 & \cite{Lorenzi_2014} & 27.9 & \cite{Singh_2012} &  0.205 & \cite{Singh_2012} 
\end{tabular}
\end{center}
\end{table}
\end{landscape}

\clearpage

\clearpage

\clearpage

For the current values of the temperature degradation coefficients, given in Table \ref{Table_1}, only one of the PV + TEG combinations can produce more electrical power than the PV alone, namely a-Si. Here, the total output power would be enhanced with about 30 \%. The other three combinations will only be able to produce additional power if the degradation of PV performance as a function of temperature, $\beta$, is significantly decreased. It is also seen that the TEG performance parameter, $\alpha$, only has a weak influence on the produced power, except when the value of $\alpha$ approaches zero.

\begin{figure}[t]
\begin{center}
\subfigure[PV]{\includegraphics[width=0.47\textwidth]{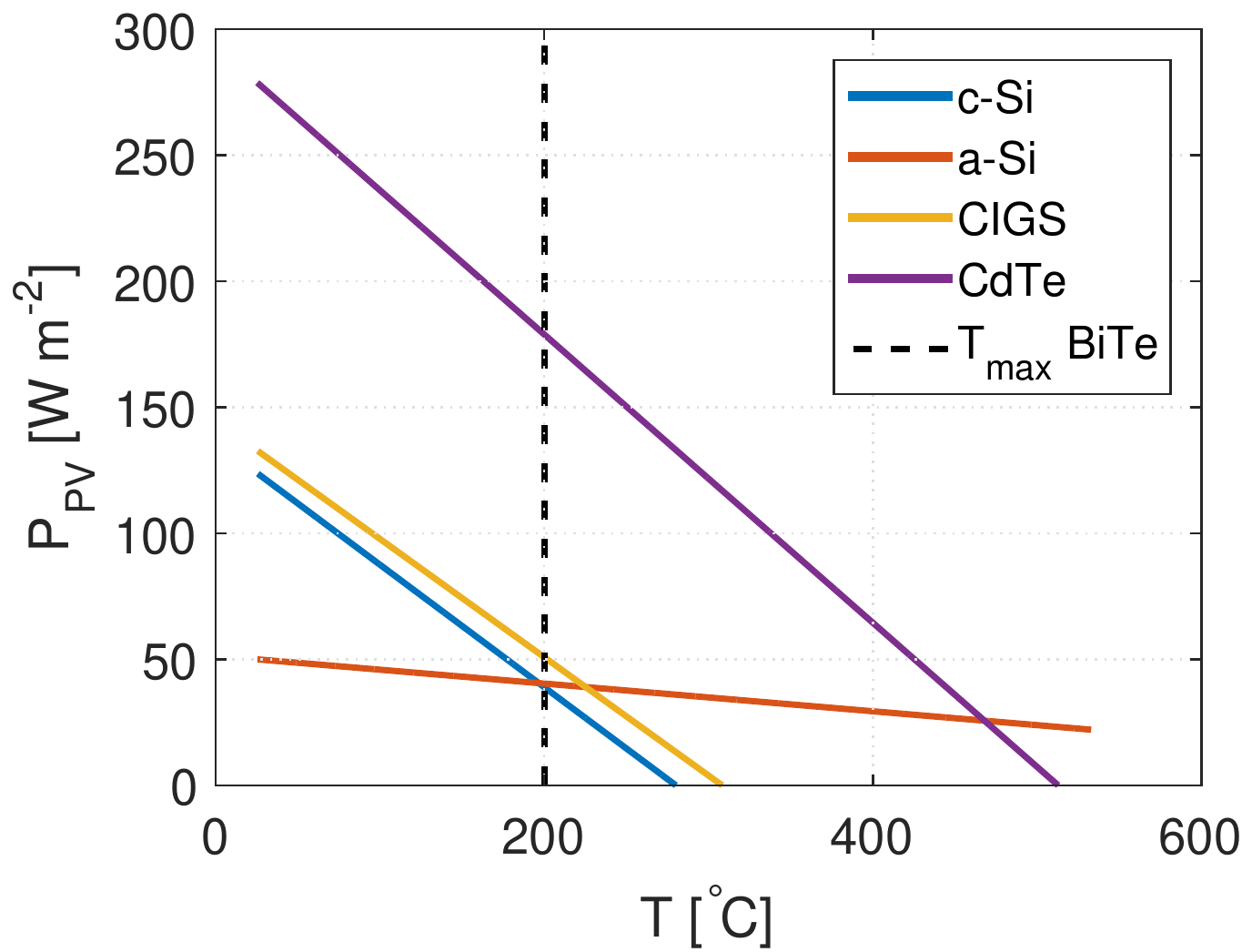}}
\subfigure[TEG]{\includegraphics[width=0.47\textwidth]{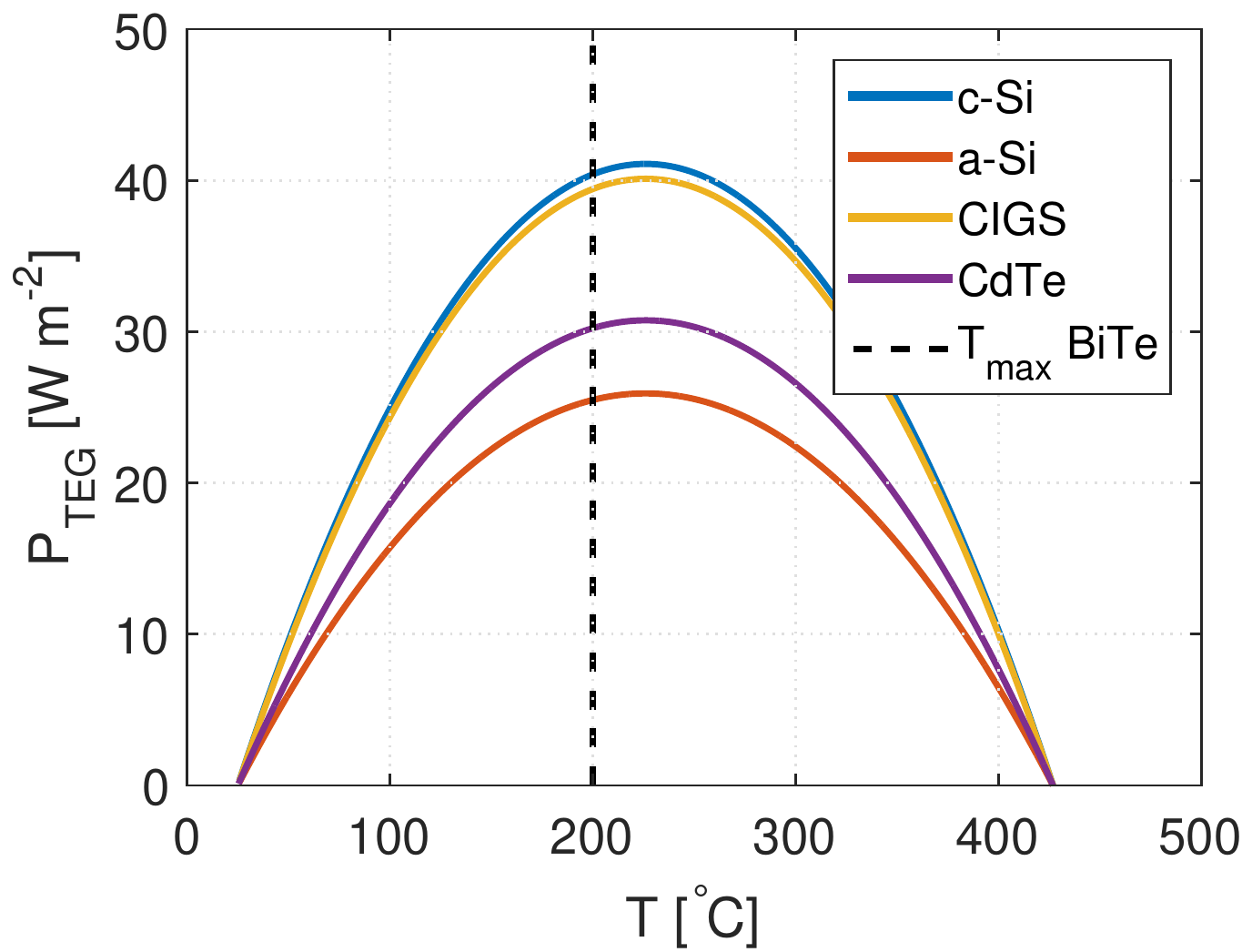}}
\end{center}
\caption{The power produced by the (a) PV and the (b) TEG, as a function of temperature. The maximum operating temperature of a BiTe TEG is indicated.}\label{Fig_PV_TEG_Power}
\end{figure}

We now consider the system with parameters as given in Table \ref{Table_1} in more detail. The power produced by the TEG and the PV, respectively, are given by the individual terms in Eq. \ref{eq_Ptot}, where the heat transported through the TEG given by Eq. \ref{eq_q_TEG}. The power is shown in Fig. \ref{Fig_PV_TEG_Power} for the case of $Q_\n{solar} = 1000$ W m$^{-2}$. As can be seen from the figure, the low efficiency of the TEG results in a lower power production than the PV. The relative fast degradation of the PV with temperature is also readily apparent. The ratio of the power produced by the PV and TEG fits well with the 10 \% previously reported \cite{Ju_2012}.

Note that the analysis is conducted up to a temperature of 500 $^\circ$C. This is an unrealistically high temperature, and it is chosen only for completeness. A BiTe TEG cannot operate above 250 $^\circ$C, as this will cause chemical decomposition of the TEG. Also, the efficiency of e.g. crystalline Si solar cells are known to drop to zero at 270 $^\circ$C\cite{Evans_1978,Skoplaki_2009}.

\begin{figure}
  \centering
  \includegraphics[width=1\columnwidth]{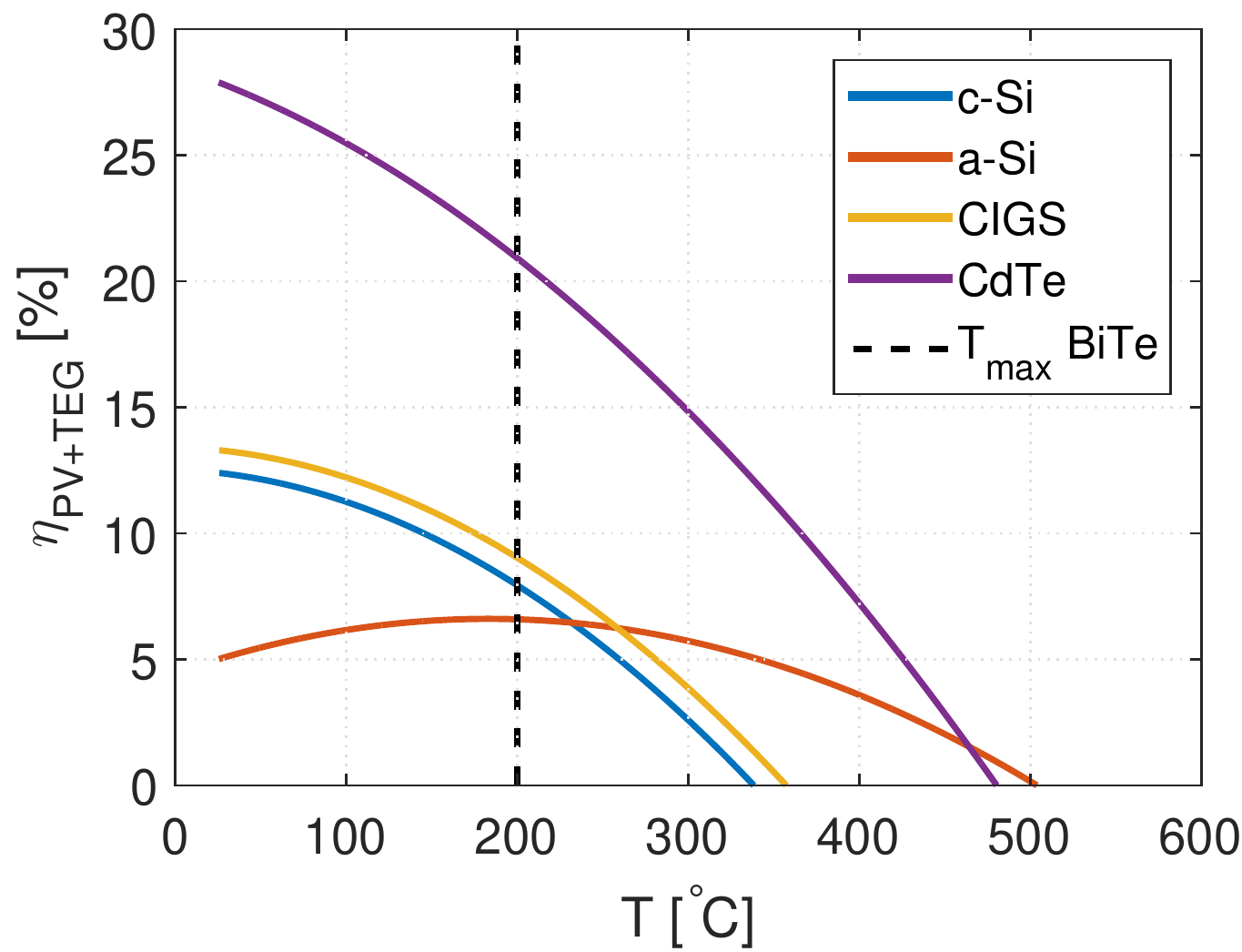}
  \caption{The efficiency of the combined PV + TEG system. The efficiency is directly proportional to the produced power, with the proportionally factor being $Q_\n{solar}$. The maximum operating temperature of a BiTe TEG is indicated.}
  \label{Fig_PV_TEG_T_eta_Comb}
\end{figure}

The efficiency of a coupled PV and TEG system are given as
\begin{equation}
\eta_\n{PV + TEG}(T)=\eta_\n{PV}(T)+\eta_\n{TEG}(T)(1-\eta_\n{PV}(T)-\eta_\n{non})
\end{equation}
This efficiency is shown in Fig. \ref{Fig_PV_TEG_T_eta_Comb} for the four different PVs considered. From the figure it is seen that only for the case of a-Si does the addition of a TEG increase the efficiency and power produced by the system. For the remaining systems the degradation in performance of the PV with increasing temperature is much greater than the power produced by the TEG.

\section{Discussion}
The results clearly demonstrate that coupling a PV and a TEG in most cases will result in a device with a poorer performance than the PV alone. This is in agreement with previous results, where an increase of performance from negative values to at most 1.3-1.8 \% compared to that of the PV alone has been observed \cite{Vorobiev_2006,Mizoshiri_2012,Najafi_2013,Lin_2014}. The work is also in agreement with that of \citet{Sark_2011}, for the realistic TEG value of $Z=0.001$ K$^{-1}$, which roughly corresponds to real-world TEG devices.  Finally, the relative efficiency of the PV + TEG system also depends on whether the PV + TEG system is compared to a PV operating at 25 $^\circ{}$C, or operating at the hot side temperature, but without the TEG. In the latter case, the combined PV + TEG system will of course always have a higher efficiency than the PV system alone, as also found by \citet{Sark_2011}.

The results determined above is in direct contrast with the class of previously discussed reported results, which have demonstrated an increase in performance of more than 50 \% from a combined PV + TEG system, even though the systems were operating at room temperature and the temperature span across the TEG was $<20$ $^\circ$C\cite{Wang_2011,Park_2013,Zhang_2013,Deng_2013,Hsueh_2014}. We cannot explain these results in the framework discussed in this article, nor from examining the known and well characterized performance of commercial TEGs or previously published (optimistic) PV + TEG modeling \cite{Sark_2011}. That the TEG itself can produce a significant amount of power with a temperature span lower than $<20$ $^\circ$C remains very implausible.

Regarding the use of combined TEG+PV systems one can envision other systems than the direct coupling discussed above. The wavelength separating approach, where the thermal ``part'' of the spectrum is directed to a TEG does not suffer the degradation in performance seen in the direct coupling, as the temperature of the PV is not increased \cite{Zhang_2005,Kraemer_2008,Mizoshiri_2012,Ju_2012}. There could also be a benefit of increasing the concentration of the solar radiation \cite{Zhang_2014}.

Also, a direct coupled device might still be interesting in applications where the amount of generated power is not critical. This could be a sensor application, where the PV can produce power during sunlight and the TEG can produce power from a temperature difference e.g. at night.

\section{Conclusion}
The combined performance of a solar photovoltaic (PV) and thermoelectric generator (TEG) system was examined for four different types of PVs and a commercially available bismuth telluride TEG. The degradation of PV performance with temperature was shown to be much faster than the increase in power produced by the TEG, due to the low efficiency of the TEG. For the cases of crystaline Si (c-Si), copper indium gallium (di)selenide (CIGS) and cadmium telluride (CdTe) PV cells the combined system produced a lower power and had a lower efficiency than the PV alone, whereas for an amorphous Si (a-Si) cell the performance could be slightly increase by the TEG. A coupled PV + TEG system is thus not a viable option for power production as long as the PV performance decreases significantly with increasing temperature.

\appendix
\section{Maximum efficiency of PV + TEG system}\label{sec_app_max_eff}
From Eqs. (\ref{eq_eff_TEG}), (\ref{eq_q_TEG}) and (\ref{eq_Ptot}) it is possible to derive the temperature at which the total power output is maximum. This is given by
\begin{eqnarray}
T_\mathrm{max}=\frac{1}{4}\frac{1}{\alpha\beta\etapv}\left(\pm2T_c\alpha\beta\etapv\pm2T_\mathrm{ref}\alpha\beta\etapv\right.\nonumber\\
\mp\beta\delta\etapv\pm2\alpha\eta_\mathrm{non}\pm2\alpha\eta_pv0\mp2\alpha\nonumber\\
\pm\left(4T_c^2\alpha^2\beta^2\etapv^2-8T_c*T_\mathrm{ref}\alpha^2\beta^2\etapv^2+4T_\mathrm{ref}^2\alpha^2\beta^2\etapv^2\right.\nonumber\\
-4T_c\alpha\beta^2\delta\etapv^2+4T_\mathrm{ref}\alpha\beta^2\delta\etapv^2\nonumber\\
-8T_c\alpha^2\beta\eta_\mathrm{non}\etapv-8T_c\alpha^2\beta\etapv^2+8T_\mathrm{ref}\alpha^2\beta\eta_\mathrm{non}\etapv\nonumber\\
+8T_\mathrm{ref}\alpha^2\beta\etapv^2+\beta^2\delta^2\etapv^2+8T_c\alpha^2\beta\etapv\nonumber\\
-8T_\mathrm{ref}\alpha^2\beta\etapv+8\alpha\beta^2\etapv^2+4\alpha\beta\delta\eta_\mathrm{non}\etapv\nonumber\\
+4\alpha\beta\delta\etapv^2+4\alpha^2\eta_\mathrm{non}^2\nonumber\\
+8\alpha^2\eta_\mathrm{non}\etapv+4\alpha^2\etapv^2-4\alpha\beta\delta\etapv\nonumber\\
\left.\left.-8\alpha^2\eta_\mathrm{non}-8\alpha^2\etapv+4\alpha^2\right)^\frac{1}{2}\right)\label{eq_Ptotmax_T}
\end{eqnarray}

The total power produced by the PV + TEG can then be found using Eq. (\ref{eq_Ptot}) and the solution for the optimum temperature given in the expression above (Eq. \ref{eq_Ptotmax_T}).


\begin{thebibliography}{}

\bibitem[Attivissimo et~al., 2015]{Attivissimo_2015}
Attivissimo, F., Di~Nisio, A., Lanzolla, A. M.~L., Paul, M., 2015. Feasibility of a photovoltaic-thermoelectric generator: performance analysis and simulation results. IEEE T. Instrum. Meas. 64, 1158-1169.

\bibitem[Deng et~al., 2013]{Deng_2013}
Deng, Y., Zhu, W., Wang, Y., Shi, Y., 2013. Enhanced performance of solar-driven photovoltaic-thermoelectric hybrid system in an integrated design. Sol. Energy 88, 182-191.

\bibitem[Dunn and Gostein, 2012]{Dunn_2012}
Dunn, L., Gostein, M., 2012. Light soaking measurements of commercially available CIGS PV modules. Photovolt. Spec. Conf. (PVSC) 2012, 1260-1265.

\bibitem[Emery et~al., 1996]{Emery_1996}
Emery, K., Burdick, J., Caiyem, Y., Dunlavy, D., Field, H., Kroposki, B., Moriarty, T., Ottoson, L., Rummel, S., Strand, T., Wanlass, M.W., 1996. Temperature dependence of photovoltaic cells, modules and systems. Photovolt. Spec. Conf. (PVSC) 1996, 1275-1278.

\bibitem[Evans and Florschuetz, 1978]{Evans_1978}
Evans, D., Florschuetz, L. 1978. Terrestrial concentrating photovoltaic power system studies. Sol. Energy 20, 37-43.

\bibitem[Fisac et~al., 2014]{Fisac_2014}
Fisac, M., Villasevil, F.~X., L{\'o}pez, A.~M., 2014. High-efficiency photovoltaic technology including thermoelectric generation. J. Power Sources 252, 264-269.

\bibitem[Hsueh et~al., 2015]{Hsueh_2014}
Hsueh, T.-J., Shieh, J.-M., Yeh, Y.-M., 2015. Hybrid Cd-free CIGS solar cell/teg device with ZnO nanowires. Prog. Photovoltaics 23, 507-512.

\bibitem[Hulk Energy Technology~Co., 2015]{Hulk_2015}
Hulk Energy Technology Co., Ltd., Taiwan 2015. Data sheet on CdF-1100E1 CIGS PV.

\bibitem[Hung et~al., 2015]{Hung_2015}
Hung, L.~T., Van~Nong, N., Han, L., Bj\o{}rk, R., Ngan, P.~H., Holgate, T.~C., Linderoth, S., Pryds, N., 2015. Segmented thermoelectric oxide-based module. Accepted for publication in Energ. Tech.

\bibitem[Ju et~al., 2012]{Ju_2012}
Ju, X., Wang, Z., Flamant, G., Li, P., Zhao, W., 2012. Numerical analysis and optimization of a spectrum splitting concentration photovoltaic-thermoelectric hybrid system. Sol. Energy 86, 1941-1954.

\bibitem[Kiflemariam et~al., 2014]{Kiflemariam_2014}
Kiflemariam, R., Almas, M., Lin, C., 2014. Modeling integrated thermoelectric generator-photovoltaic thermal (TEG-PVT) system. Proc. 2014 COMSOL Conf.

\bibitem[Kim et~al., 2012]{Kim_2012}
Kim, C., Kim, D.~H., Kim, H., Chung, J.~S., 2012. Significant enhancement in the thermoelectric performance of a bismuth telluride nanocompound through brief fabrication procedures. ACS Appl. Mater. Interfaces 4, 2949-2954.

\bibitem[Kraemer et~al., 2008]{Kraemer_2008}
Kraemer, D., Hu, L., Muto, A., Chen, X., Chen, G., Chiesa, M., 2008. Photovoltaic-thermoelectric hybrid systems: A general optimization methodology. Appl. Phys. Lett. 92, 243503.

\bibitem[Liao et~al., 2014]{Liao_2014}
Liao, T., Lin, B., Yang, Z., 2014. Performance characteristics of a low concentrated photovoltaic--thermoelectric hybrid power generation device. Int. J. Therm. Sci. 77, 158-164.

\bibitem[Lin et~al., 2014]{Lin_2014}
Lin, W., Shih, T.-M., Zheng, J.-C., Zhang, Y., Chen, J., 2014. Coupling of temperatures and power outputs in hybrid photovoltaic and thermoelectric modules. Int. J. Heat Mass Tran. 74, 121-127.

\bibitem[Lorenzi et~al., 2014]{Lorenzi_2014}
Lorenzi, B., Acciarri, M., Narducci, D., 2014. Analysis of thermal losses for a variety of single-junction photovoltaic cells: An interesting means of thermoelectric heat recovery. J. Electron. Mater. 44, 1809-1813.

\bibitem[Luque and Mart{\i}, 1999]{Luque_1999}
Luque, A., Mart{\i}, A., 1999. Limiting efficiency of coupled thermal and photovoltaic converters. Sol. Energ. Mat. Sol. C. 58, 147-165.

\bibitem[Ma et~al., 2008]{Ma_2008}
Ma, Y., Hao, Q., Poudel, B., Lan, Y., Yu, B., Wang, D., Chen, G., Ren, Z., 2008. Enhanced thermoelectric figure-of-merit in p-type nanostructured bismuth antimony tellurium alloys made from elemental chunks. Nano Lett. 8, 2580-2584.

\bibitem[Makrides et~al., 2009]{Makrides_2009}
Makrides, G., Zinsser, B., Georghiou, G.~E., Schubert, M., Werner, J.~H., 2009. Temperature behaviour of different photovoltaic systems installed in Cyprus and Germany. Sol. Energ. Mat. Sol. C. 93, 1095-1099.

\bibitem[Makrides et~al., 2012]{Makrides_2012}
Makrides, G., Zinsser, B., Norton, M., Georghiou, G.~E., 2012. Performance of photovoltaics under actual operating conditions. INTECH Open Access Publisher.

\bibitem[Marlow~Industries, 2014]{Marlow_2014}
Marlow Industries, Inc, 2014. Datasheet on TG12-series

\bibitem[Micropelt, 2015]{Micropelt}
Micropelt, 2015. Datasheet.

\bibitem[Mizoshiri et~al., 2012]{Mizoshiri_2012}
Mizoshiri, M., Mikami, M., Ozaki, K., 2012. Thermal-photovoltaic hybrid solar generator using thin-film thermoelectric modules. Jpn. J. Appl. Phys. 51, 06FL07.

\bibitem[Mohring and Stellbogen, 2008]{Mohring_2008}
Mohring, H., Stellbogen, D., 2008. Annual energy harvest of pv systems--advantages and drawbacks of different PV technologies. 23rd Eur. Photovolt. Sol. Energ. Conf., 1-5.

\bibitem[Najafi and Woodbury, 2013]{Najafi_2013}
Najafi, H., Woodbury, K.~A., 2013. Modeling and analysis of a combined photovoltaic-thermoelectric power generation system. J. Sol. Energ. 135, 031013.

\bibitem[Park et~al., 2013]{Park_2013}
Park, K.-T., Shin, S.-M., Tazebay, A.~S., Um, H.-D., Jung, J.-Y., Jee, S.-W.,
  Oh, M.-W., Park, S.-D., Yoo, B., Yu, C., J.-H. Lee., 2013. Lossless hybridization between photovoltaic and thermoelectric devices.  Sci. rep. 3, 2123.

\bibitem[Santbergen and van Zolingen, 2008]{Santbergen_2008}
Santbergen, R., van Zolingen, R.~C., 2008. The absorption factor of crystalline silicon pv cells: a numerical and experimental study. Sol. Energ. Mat. Sol. C. 92, 432-444.

\bibitem[Singh and Ravindra, 2012]{Singh_2012}
Singh, P., Ravindra, N., 2012. Temperature dependence of solar cell performance — an analysis. Sol. Energ. Mat. Sol. C. 101, 36-45.

\bibitem[Skoplaki and Palyvos, 2009]{Skoplaki_2009}
Skoplaki, E., Palyvos, J., 2009. On the temperature dependence of photovoltaic module electrical performance: A review of efficiency/power correlations. Sol. Energy 83, 614-624.

\bibitem[Solibro~GmbH, 2015]{Solibro_2015}
Solibro GmbH, Germany, 2015. Datasheet on SL2.

\bibitem[Sundarraj et~al., 2014]{Sundarraj_2014}
Sundarraj, P., Maity, D., Roy, S.~S., Taylor, R.~A., 2014. Recent advances in thermoelectric materials and solar thermoelectric generators - a critical review. RSC Adv. 4, 46860-46874.

\bibitem[TSMC Solar Europe~GmbH, 2015]{TSMC_2015}
TSMC Solar Europe GmbH, Germany, 2015. Datasheet on TS-140C CIGS.

\bibitem[Van~Sark, 2011]{Sark_2011}
Van~Sark, W., 2011. Feasibility of photovoltaic--thermoelectric hybrid modules, Appl. Energy 88, 2785-2790.

\bibitem[Vorobiev et~al., 2006]{Vorobiev_2006}
Vorobiev, Y., Gonz{\'a}lez-Hern{\'a}ndez, J., Vorobiev, P., Bulat, L., 2006. Thermal-photovoltaic solar hybrid system for efficient solar energy conversion. Sol. Energy 80, 170-176.

\bibitem[Wang et~al., 2011]{Wang_2011}
Wang, N., Han, L., He, H., Park, N.-H., and Koumoto, K. (2011). A novel high-performance photovoltaic-thermoelectric hybrid device. Energ. Environ. Sci. 4, 3676-3679.

\bibitem[Wu et~al., 2015]{Wu_2015}
Wu, Y.-Y., Wu, S.-Y., and Xiao, L. (2015). Performance analysis of photovoltaic--thermoelectric hybrid system with and without glass cover. Energ. Convers. Manage. 93, 151-159.

\bibitem[Xu et~al., 2014]{Xu_2014}
Xu, X., Zhou, S., Meyers, M.~M., Sammakia, B.~G., Murray, B.~T., 2014. Performance analysis of a combination system of concentrating photovoltaic/thermal collector and thermoelectric generators. J. Electron. Packaging 136, 041004.

\bibitem[Zhang et~al., 2014]{Zhang_2014}
Zhang, J., Xuan, Y., Yang, L., 2014. Performance estimation of photovoltaic-thermoelectric hybrid systems. Energy 78, 895-903.

\bibitem[Zhang et~al., 2005]{Zhang_2005}
Zhang, Q.~J., Tang, X.~F., Zhai, P.~C., Niino, M., Endo, C., 2005. Recent development in nano and graded thermoelectric materials. Mater. Sci. Forum 492, 135-140.

\bibitem[Zhang et~al., 2013]{Zhang_2013}
Zhang, Y., Fang, J., He, C., Yan, H., Wei, Z., Li, Y., 2013. Integrated energy-harvesting system by combining the advantages of polymer solar cells and thermoelectric devices. J. Phys. Chem. C 117, 24685-24691.

\end{thebibliography}
\end{document}